\documentclass[sigconf,authorversion]{acmart}

\AtBeginDocument{%
  }

\copyrightyear{2023} 
\acmYear{2023} 
\setcopyright{acmlicensed}\acmConference[CIKM '23]{Proceedings of the 32nd ACM International Conference on Information and Knowledge Management}{October 21--25, 2023}{Birmingham, United Kingdom}
\acmBooktitle{Proceedings of the 32nd ACM International Conference on Information and Knowledge Management (CIKM '23), October 21--25, 2023, Birmingham, United Kingdom}
\acmPrice{15.00}
\acmDOI{10.1145/3583780.3614963}
\acmISBN{979-8-4007-0124-5/23/10}
\begin{document}

\title{MemoNet: Memorizing All Cross Features' Representations Efficiently via Multi-Hash Codebook Network for CTR Prediction}
\renewcommand{\shorttitle}{MemoNet: Memorizing All Cross Features' Representations Efficiently via Multi-Hash Codebook Network\\for CTR Prediction}

\author{Pengtao Zhang}
\affiliation{%
  \institution{Sina Weibo}
   \city{Beijing}
   \country{China}
}
\email{zpt1986@126.com}

\author{Junlin Zhang}
\affiliation{%
  \institution{Sina Weibo}
   \city{Beijing}
   \country{China}
}
\email{junlin6@staff.weibo.com}

\renewcommand{\shortauthors}{Pengtao Zhang and Junlin Zhang}

\begin{abstract}
 New findings in natural language processing (NLP) demonstrate that the strong memorization capability contributes a lot to the success of Large Language Models (LLM). This inspires us to  explicitly bring an independent memory mechanism into CTR ranking model to learn and memorize cross features' representations. In this paper, we propose multi-Hash Codebook NETwork (HCNet) as the memory mechanism for efficiently learning and memorizing representations of cross features in CTR tasks. HCNet uses a multi-hash codebook as the main memory place and the whole memory procedure consists of three phases: multi-hash addressing, memory restoring, and feature shrinking. We also propose a new CTR model named MemoNet which combines HCNet with a DNN backbone. Extensive experimental results on three public datasets and online test show that MemoNet reaches superior performance over state-of-the-art approaches. Besides, MemoNet shows scaling law of large language model in NLP, which means we can enlarge the size of the codebook in HCNet to sustainably obtain performance gains. Our work demonstrates the importance and feasibility of learning and memorizing representations of cross features, which sheds light on a new promising research direction. The source code is in \url{https://github.com/ptzhangAlg/RecAlg}.
  
\end{abstract}






\maketitle

\section{INTRODUCTION}

Click-through rate (CTR) prediction plays an important role 
\begin{figure}[h]
  \centering
  \includegraphics[width=1.0\linewidth]{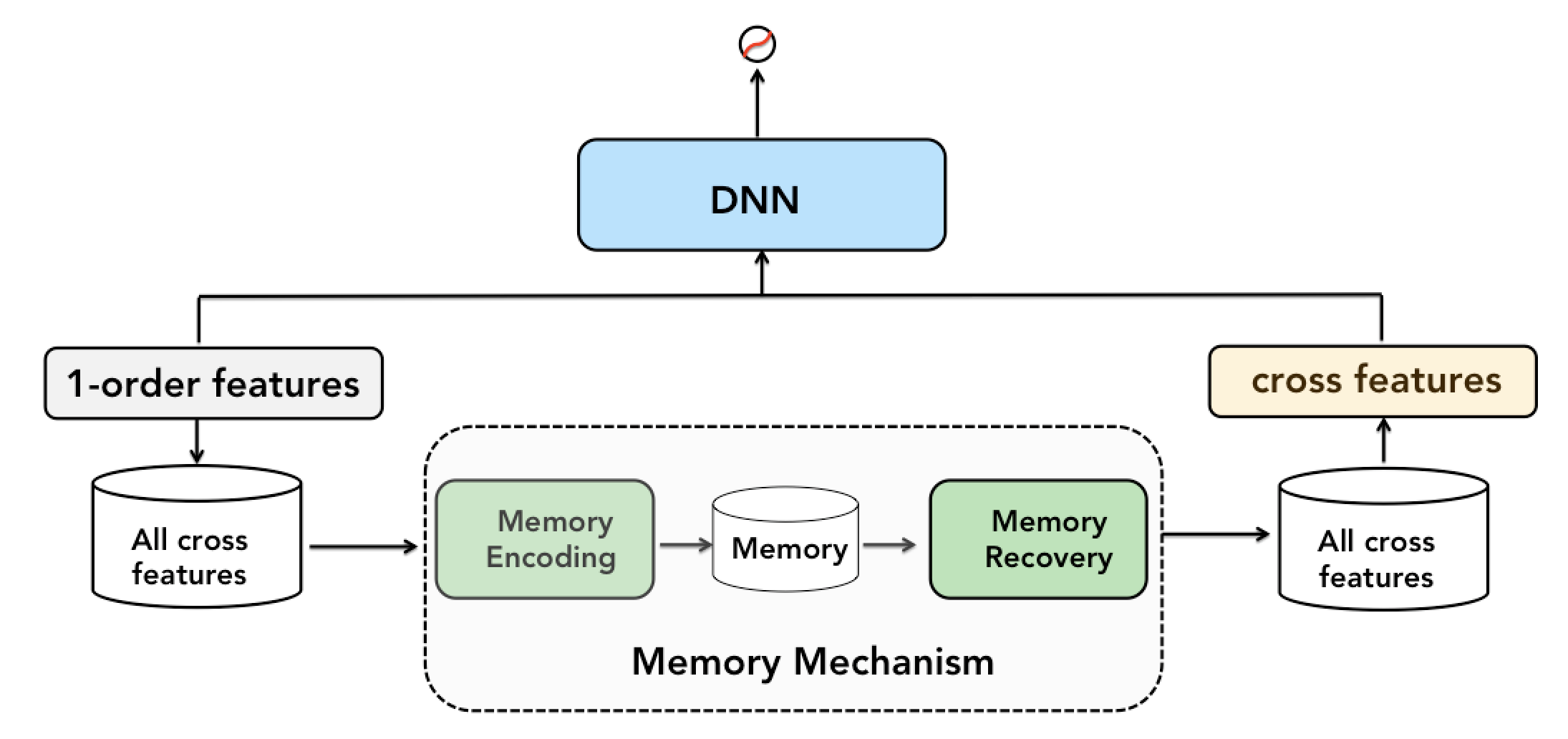}
  \caption{Learning and memorizing representations of all cross features through independent memory mechanism}
  \Description{}
  \label{Fig.framework}
\end{figure}
in personalized advertising and recommender system, and it aims to 
predict the probability of a user clicking on the recommended items. 
Many models\cite{rendle2010factorization,pan2018field,nfm,xiao2017attentional,guo2017deepfm,wang2017deep,masknet} have been proposed to resolve this problem. However, researches\cite{HuangZZ19,lian2018xdeepfm,WangSCJLHC21} show that increasing model parameters of modern deep CTR models has little influence on model performance, which is contrary to the phenomenon shown in lots of large language models\cite{gpt4,xlnet,gpt3} in NLP. The past several years have witnessed the great success of Large Language Models\cite{xlnet,bert,gpt2,gpt3,t5,gpt4} in natural language processing with ever larger model size such as Bert\cite{bert}, GPT-2\cite{gpt2}, T5\cite{t5}, GPT-3\cite{gpt3} and GPT-4\cite{gpt4}. Given ample training data, these big models show a great advantage that models can continually absorb linguistic and world knowledge from data and model performance can be boosted constantly by increasing the number of model parameters, which is named "scaling law" and is always regarded as the most prominent feature of big models. 

In order to explain the success of large language models, many recent works\cite{carlini2022quantifying,tirumala2022memorization,zhu2020modifying,webson2021prompt} discuss the memorization and generalization ability of these models. Some works\cite{heinzerling2020language,wang2021language,magar2022data} show that big models in NLP have great capabilities to memorize factual world knowledge in their vast amount of parameters during training, which is a crucial component of their improved generalization.

Useful cross feature is almost the most important knowledge contained in data of CTR tasks. The fact that larger CTR model doesn't lead to better model performance suggests current network structure is inefficient to extract and memorize knowledge such as cross features in the data. To tackle this problem and inspired by new findings in NLP,  we aim to mimic the strong memory capability of big models in NLP by explicitly introducing an independent memory mechanism into CTR ranking model to learn and memorize cross features' representations. Meanwhile, we also care about the following question: \textbf{Can we boost model performance constantly by enlarging the size of this memory just as scaling law shown by big models in NLP? }

Our proposed conceptual independent memory mechanism is shown in Figure 1, which is composed of three modules: the "memory module" is the main place to store cross features' representations and the "memory encoding module" is used to encode the representation, while the "memory recovery module" tries to retrieve each cross feature's representation for use. 

We argue in this paper that an independent memory mechanism for learning and memorizing representations of cross features reflects the thought of a clear division of labour for each component in modern DNN ranking model. The memory system is in charge of extracting and memorizing knowledge in cross features and the DNN part is only responsible for the generalization of all features, including 1-order, 2-order, and high order features, which greatly reduces DNN part's learning burden and boosts model performance. It is worth noting that, when comparing with other SOTA CTR approaches which explicitly model feature interactions through sub-network such as xDeepFM\cite{lian2018xdeepfm},  DCN v2\cite{WangSCJLHC21}, CAN\cite{zhou2020can} and FiBiNet\cite{HuangZZ19}, learning and memorizing cross features through an independent memory mechanism instead of modeling cross features by a special DNN sub-network shows a new and different research direction. Besides, though a baseline method named "cartesian product" is introduced in CAN\cite{zhou2020can}, which allocates one embedding for each cross feature, it's not feasible in real-life scenario because of the combination explosion problem.

To make this idea more specific, in this paper we propose multi-Hash Codebook NETwork(HCNet) as the above-mentioned memory mechanism for learning and memorizing representations of cross features in CTR tasks. In HCNet, multi-hash codebook is used as the place to store cross features' representations, which consists of a large amount of codewords as the basic memory unit.  We want to use as little codeword as possible while exactly memorizing each cross feature's information at the same time. Therefore, it's unavoidable for one codeword to store a large amount of cross features' information and we aim to restore the exact representation for each cross feature from it. The whole procedure looks like information encryption and decryption, that's the reason why we name the basic memory unit "codeword" and the main memory place "codebook".

For better representation, we use multi-hash functions to segment the representation of cross feature into  $m$ chunks and each chunk represents part of the complete representation.  In order to efficiently store and retrieve representation of any feature interaction, the entire procedure of HCNet could be divided into three phases: multi-hash addressing, memory restoring, and feature shrinking. Multi-hash addressing phase aims to obtain all cross features in an input instance and locate addresses of all chunks in the codebook for each cross feature. Memory restoring stage recovers the exact representation for any cross feature through "linear memory restoring" or "attentive memory restoring" method, and feature shrinking phase aims to compress representations of many cross features in input instance into a thin embedding layer to reduce the model parameters. Though multi-hash codebook is rather concise memory and more complex structure can be used, we aim to verify the feasibility of learning and memorizing representations of all cross features with limited resources and we will leave more complex structure as future work.

As a general module to efficiently memorize cross features, HCNet can be incorporated into any current deep CTR models. DNN model is the simplest deep CTR model and we use it as a backbone to combine it with HCNet to boost the model performance, which is  called  "MemoNet" in this paper.

Extensive experiments on three real-world public datasets and online test show that MemoNet leads to significant performance improvements compared to state-of-the-art CTR methods. Besides, MemoNet shows scaling law of big models, that is to say, the model can absorb more knowledge from data when we increase model parameters through memorizing. This demonstrates we can just enlarge the codeword number of codebook in HCNet to obtain performance gains sustainably. As far as we know, MemoNet is the first CTR model with this kind of ability.

The main contributions of our works are concluded as follows:

(1) To the best of our knowledge, we are the first to propose and verify the feasibility of learning and memorizing all cross features  with limited resources, which sheds light on a new promising research direction.

(2) We propose a novel HCNet as a specific memory mechanism and it's a general module to be incorporated into any current
deep CTR model to greatly boost model performance.

(3) We plug HCNet into DNN models to form "MemoNet", which is the first CTR model to show the scaling law of big models as large language models in NLP show.

(4) The significant improvements on three real-world benchmarks and online tests confirm the effectiveness of HCNet in CTR ranking systems.
\section{Related Works}
Many deep learning based CTR models\cite{guo2017deepfm, wang2017deep, WangSCJLHC21,song2019autoint,HuangZZ19,lian2018xdeepfm,masknet} have been proposed in recent years and how to effectively model feature interactions is key to them.
\subsection{Deep CTR Methods Modeling Feature Interactions}
While early stage deep CTR models process feature interactions in implicit way\cite{zhang2016deep, xiao2017attentional, nfm}, most recent works explicitly model pairwise or high order feature interactions by sub-network\cite{guo2017deepfm, lian2018xdeepfm, wang2017deep,WangSCJLHC21,HuangZZ19,song2019autoint,zhou2020can}. For example, DeepFM\cite{guo2017deepfm} utilizes FM to capture 2-order feature interactions. Deep \& Cross Network (DCN)\cite{wang2017deep} and DCN V2 \cite{WangSCJLHC21} efficiently capture feature interactions of bounded degrees in an explicit fashion. Similarly, xDeepFM \cite{lian2018xdeepfm} models high-order feature interactions by proposing a novel Compressed Interaction Network (CIN). AutoInt\cite{song2019autoint} designs a multi-head self-attentive neural network with residual connections to explicitly model the feature interactions. FiBiNet \cite{HuangZZ19} and FiBiNet++\cite{fibnetjia} introduce bilinear feature interaction function while MaskNet \cite{masknet} proposes instance-guided mask to capture high order cross features. As an extreme example to capture cross feature, CAN \cite{zhou2020can} proposes a Co-Action Network to dynamically learn a mico-MLP sub-network for each feature to fit complex feature interactions, which leads to a large amount of model parameters. Many reported results \cite{lian2018xdeepfm,wang2017deep,WangSCJLHC21,HuangZZ19} show that explicitly modeling high-order interaction  outperforms approaches in implicit ways. 

However, majority of real-life CTR ranking systems still need  humanly designed cross features for better performance, which demonstrates inefficiency of deep networks modeling feature interactions explicitly. An independent memory mechanism for memorizing representation of cross features helps MLP focus on feature generalization instead of modeling feature interaction, which make learning easier. We will verify this in the experiment part that memorizing is a much more efficient way to learn cross features.

\subsection{Deep CTR Methods Mining Feature Interactions}

Another line of research on pairwise and high order features  aims to mine some useful feature interactions in multiple stages: a small proportion of feature interactions are mined by manual feature selection or AutoML in the first stage and are introduced into CTR model as new features in the second stage. For example, Wide \& Deep Learning\cite{cheng2016wide} jointly trains wide linear models and deep neural networks to combine the benefits of memorization and generalization. However, expertise feature engineering is still needed. To alleviate manual efforts in feature engineering, AutoFIS\cite{autofis020} and AutoGroup\cite{autogroup20} use AutoML to seek a small proportion of useful high-order feature interactions to train CTR model in two or three stages.

There are two shortcomings in these multi-stage mining methods: First, most real-world ranking systems support online learning, and one difficulty of these approaches is that they can't instantly capture new beneficial feature interactions. In addition, these models miss a large amount of long-tail useful cross features, which have a great impact on model performance  as a whole. Different from existing studies, our proposed approach can be deployed in online learning systems and capture long-tail beneficial cross features efficiently. 

\section{PRELIMINARIES}
DNN model is almost the simplest one among deep models and is always used as a sub-component in most current DNN ranking systems\cite{guo2017deepfm,wang2017deep,WangSCJLHC21,lian2018xdeepfm,HuangZZ19,cheng2016wide}. It contains three components: feature embedding, MLP and prediction layer. 

As for the feature embedding, we follow other works\cite{guo2017deepfm,wang2017deep,WangSCJLHC21,lian2018xdeepfm,HuangZZ19,cheng2016wide} to map one-hot representation to dense, low-dimensional embedding vectors suitable for complex transformation. We can obtain feature embedding $\mathbf{v}_i$ for one-hot vector $\mathbf{x}_i$ via:
  \begin{equation}
    \mathbf{v}_i = \mathbf{x}_i\mathbf{W}_e
  \end{equation}
where $\mathbf{W}_e \in\mathbb{R}^{d\times n}$ is the embedding matrix of $n$ features and $d$ is the dimension of field embedding.

\begin{figure*}[h]
  \centering
  \includegraphics[width=0.8\linewidth]{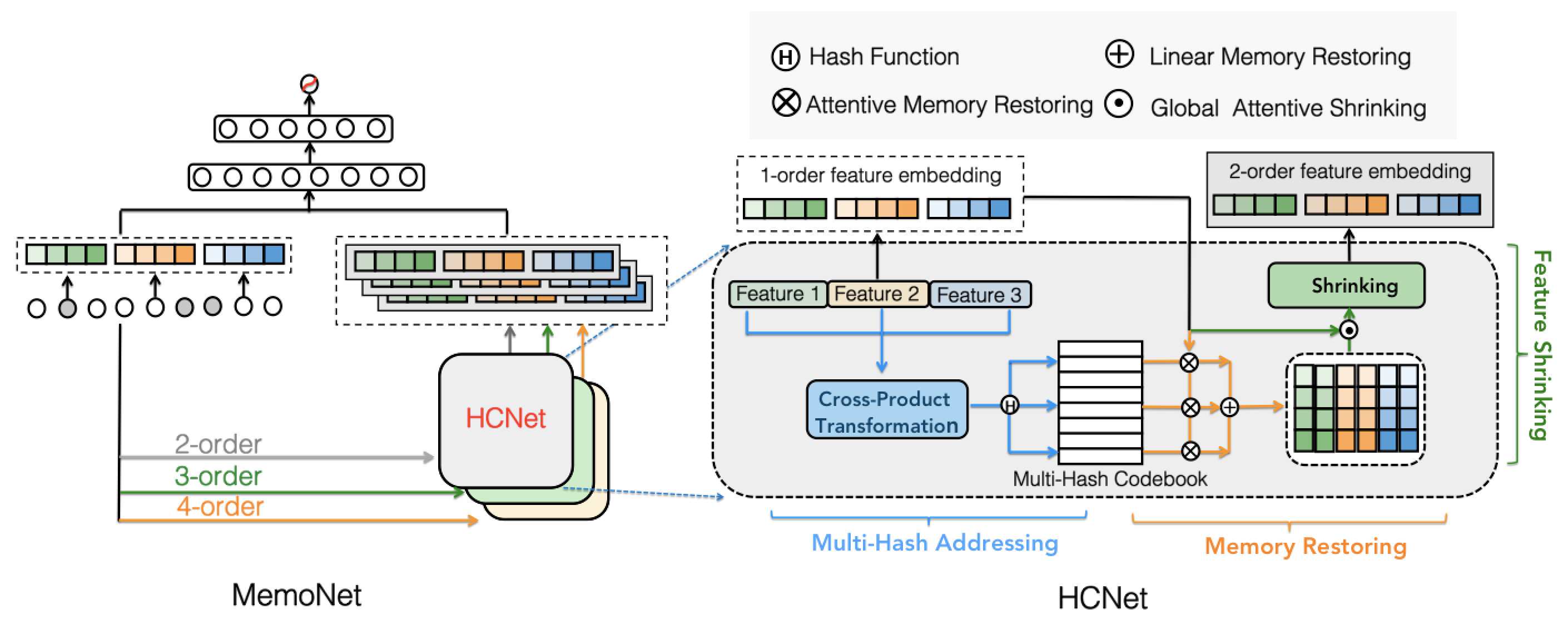}
  \caption{The overall architectures of MemoNet and HCNet}
  \Description{}
\end{figure*}

To learn high-order feature interactions, multiple feed-forward layers are stacked on the concatenation of dense features represented as $\mathbf{H}_0 = concat\{\mathbf{v}_{1},\mathbf{v}_{2},...,\mathbf{v}_{f}\}$. Then, the feed forward process of MLP is:
\begin{equation}
  \mathbf{H}_l = ReLU(\mathbf{W}_l\mathbf{H}_{l-1} + \beta_l)
\end{equation}
where $l$ is the depth and ReLU is the activation function. $\mathbf{W}_l, \beta_l, \mathbf{H}_l$ are weighting matrix, bias and output of the $l$-th layer. 

The prediction layer is put on the last layer of multiple feed-forward networks, and the model’s output is:
\begin{equation}
  \hat{y} = \delta(\mathbf{w}_0 + \sum^n_{i=1}\mathbf{w}_i\mathbf{x}_i)
\end{equation}
where $\hat{y} \in (0, 1)$ is the predicted value of CTR, $\delta\left(\cdot\right)$ is the sigmoid function, $n$ is the size of feed-forward layer, $\mathbf{x}_i$ is the bit value of feed-forward layer and $\mathbf{w}_i$ is the learned weight for each bit value.

For binary classifications, the loss function of CTR prediction is the log loss:
\begin{equation}
  \mathcal{L} = -\frac{1}{N}\sum^N_{i=1}y_i\log(\hat{y}_i)+(1-y_i)\log(1-\hat{y}_i) 
\end{equation}
where $N$ is the total number of training instances, $y_i$ is the ground truth of $i$-th instance and $\hat{y}_i$ is the predicted CTR.  

\section{Methodology}
In this section, we first describe the multi-hash codebook in detail, which is the place to memorize representations of feature interactions. Then, we take the 2-order pairwise interaction as an example to show the technical details of
the proposed multi-Hash Codebook NETwork(HCNet) for memorizing  cross features efficiently. After that, we demonstrate our proposed MemoNet which embeds HCNet into DNN backbone for CTR prediction. Finally, we discuss how to identify key interaction fields to reduce the number of cross features.

\subsection{Multi-Hash Codebook}
Our work aims to learn and memorize all feature interactions with limited memory resources and the codebook is the place where we store cross feature's representation. Codebook consists of a large amount of codewords  which  is the basic memory unit. Formally, we define codebook as parameter matrix $\mathbf{C}\in \mathbb{R}^{n \times l}$ with size of $\mathbf{n}$ rows and $\mathbf{l}$ columns to be learned during model training. Each row of matrix $\mathbf{C}$ is one codeword which is a vector with size of $\mathbf{l}$. We leverage the row number of matrix $\mathbf{C}$ as the index or address of the corresponding codeword in the codebook.

For a feature interaction $\mathbf{x}^{(i,j)}$ coming from cross of feature $\mathbf{x}_i$ and $\mathbf{x}_j$, we can allocate a random codeword with address $\mathbf{a}^{(i,j)}$ for it through a general hash function $\mathbf{H(\cdot)}$ which projects input to a large number as follows:
\begin{equation}
  \mathbf{a}^{(i,j)}=
  \mathbf{H(\mathbf{x}^{(i,j)})} \quad MOD \quad \Vert \mathbf{C} \Vert
\end{equation}
where $\Vert \mathbf{C} \Vert=n$ means size of codebook and MOD is a remainder operator. We can use codeword embedding to memorize the cross feature $\mathbf{x}^{(i,j)}$.

Because of the combinatory explosion problems for cross features, 100 thousands or even 1 million cross features may be projected randomly in one codeword, which leads codeword embedding to a mixed representation and it's hard to distinguish them from each other. To alleviate this problem, we leverage multiple hash functions to encode cross features as follows:
\begin{equation}
  \mathbf{A}^{(i,j)} = \left\{ \mathbf{H}_t(\mathbf{x}^{(i,j)})  \quad MOD \quad \Vert \mathbf{C} \Vert\right\}_{t=1}^{m}
\end{equation}
where $H_t(\cdot)$ denotes the $t-th$ hash function and $\mathbf{A}^{(i,j)}$ is an index set indicating $m$ codeword addresses in codebook for feature $\mathbf{x}^{(i,j)}$.

The codeword embedding matrix $\mathbf{C}^{(i,j)}\in \mathbb{R}^{m \times l}$ of feature $\mathbf{x}^{(i,j)}$ can be extracted according to location index set $\mathbf{A}^{(i,j)}$ as follows:
 \begin{equation}
         \mathbf{C}^{(i,j)} = [\mathbf{C}_1^{(i,j)},\mathbf{C}_2^{(i,j)},......,\mathbf{C}_m^{(i,j)}]\in \mathbb{R}^{m \times l}
 \end{equation}
where $\mathbf{C}_t^{(i,j)}\in \mathbb{R}^{1 \times l}$ denotes the codeword embedding indicated by index produced by the $t$-th hash function $\mathbf{H}_t(\mathbf{x}^{(i,j)}) $ and $l $ is embedding size of codeword. We call the codebook with multiple hash functions "multi-hash codebook" in this paper, which is the main memory component of HCNet.

Though codeword seems similar to bin or bucket\cite{bins} in form, there are two main differences: First, bin usually doesn't  contain  so large amount of collision items like codeword does, that will bring serious damage to model performance; Second, all items in one bin will share same representation in applications while we try to recover the exact representation of one specific item in our works, and that's the most challenging task.

It's obvious that the "multi-hash codebook" segments the representation of cross feature into  $m$ chunks and each chunk indicated by a different hash function represents part of the complete information. This helps distinguish cross features from each other in the same codeword because the probability of sharing the same representation exponentially decreases with the increase of hash function number.
\subsection{HCNet}
The overall framework of HCNet is shown in Figure 2 and the entire procedure of HCNet could be divided into three phases: multi-hash addressing, memory restoring and feature shrinking.  

\textbf{Multi-Hash Addressing.}
This stage is designed to obtain all cross features in an input instance and locate addresses of all chunks in the codebook for each cross feature. Specifically, we first produce all cross features in input instance via cross product transformation. Suppose there are $f$ fields in input instance, we obtain a new feature set $\mathcal{E}_2^{all}$ with size $f\times (f-1)/2$  after cross product transformation for this instance:
 \begin{equation}
  \mathcal{E}_2^{all} = \bigcup_{i=1}^{f}\mathcal{E}_2^{i}  , \quad where \quad \mathcal{E}_2^{i} = \left\{  \mathbf{x}^{(i,j)} \right\}_{1 \leqslant j \leqslant f \quad \wedge \quad i\neq j}
\end{equation}
Here $\mathbf{x}^{(i,j)}$ means a new 2-order feature by cross of feature $\mathbf{x}_i$ and $\mathbf{x}_j$, and $\mathcal{E}_2^{i}$ is the feature set with size $(f-1)$ which contains all cross features  derived from  $\mathbf{x}_i$.

Then, we produce a global unique ID for each new feature $\mathbf{x}^{(i,j)}$ to distinguish them from each other. For feature $\mathbf{x}_i$ and $\mathbf{x}_j$, we concatenate two unique internal index numbers as a unique string type ID for new feature $\mathbf{x}^{(i,j)}$ as follows:
\begin{equation}
  \mathbf{ID}_{2}^{(i,j)} = \mathbf{ID}_2^{\left(j,i\right)} = concat[\mathbf{ID}_i,\mathbf{ID}_j]
\end{equation}
where cross feature $\mathbf{x}^{(i,j)}$ and $\mathbf{x}^{(j,i)}$ share the same ID because the input order doesn't matter for CTR prediction. If  $\mathbf{x}_i \in \mathbb{R}$ is a numerical feature, the unique $\mathbf{ID}_i$  is formed by $\mathbf{ID}_i=concat[fieldID,\phi_k(\mathbf{x}_i)]$ , where $fieldID$ is internal index number of the feature field which $\mathbf{x}_i$ belongs to and $\phi_k(\cdot)$ is a truncation function keeping $k$ decimal places for float number. We concatenate two numbers for the uniqueness of the feature ID. We find size of $k$ has an influence on model performance and the optimal setting is $k=5$.

After obtaining a unique $\mathbf{ID}_2^{(i,j)}$ for a 2-order new feature $\mathbf{x}^{(i,j)}$, we leverage hash function  $H(\cdot)$ to locate its codeword addresses in codebook as follows:
\begin{equation}
  \mathbf{A}_2^{(i,j)} = \left\{ \mathbf{H}_t(\mathbf{ID}_2^{(i,j)}) \quad MOD \quad \Vert \mathbf{C} \Vert\ \right\}_{t=1}^{m}
\end{equation}

In this way, we can locate the codeword embeddings matrix $\mathbf{C}^{(i,j)}$ for any pairwise feature $\mathbf{x}^{(i,j)}$. HCNet tunes the codeword embedding while training to learn the representation and readout the codeword embedding via multi-hash addressing in online reference stage.

\textbf{Memory Restoring.}
While "multi-hash codebook" helps distinguish cross features from each other in same codeword, it's not good enough because each chunk is still a mixed representation. To better recover each cross feature's embedding, we propose two methods in this paper.

A straightforward way to recover feature $\mathbf{x}^{(i,j)}$'s accurate representation is first to flatten the two dimensional codeword embedding matrix $\mathbf{C}^{(i,j)}\in \mathbb{R}^{m \times l}$ to one dimensional vector $\mathbf{e}^{(i,j)}$ with size of $1 \times ml $, and then project it into a new embedding space by a MLP layer:
\begin{equation}
        \mathbf{v}^{(i,j)} = \phi_1(\mathbf{e}^{(i,j)}\mathbf{W}_{1})\in \mathbb{R}^{ 1\times d}
 \end{equation}
where  $\mathbf{W}_1 \in \mathbb{R}^{ ml\times d} $ is parameter matrix of MLP layer and $\phi_1(\cdot)$ is an identity function without non-linear transformation. The new mapping based on chunks of codeword embedding extracts useful information encoded in chunks and aggregates a unique representation for cross feature. We find the non-linear transformation $\phi(\cdot)$ brings a negative effect on model performance because too early complex mutual interaction among codeword embedding chunks weakens informative signals contained in each codeword. 
Cross feature's embedding can be restored in this way and we call this method "Linear Memory Restoring (LMR)". Note we set size of the recovered cross feature  $d $, which is the same as the size of 1-order feature embedding.

In order to better restore the memorized embedding for cross feature $\mathbf{x}^{(i,j)}$, another proposed method uses  1-order embedding of feature $\mathbf{x}_i$ and $\mathbf{x}_j$ to guild the representation recovery. We think this will make memorizing and learning easier. Therefore, we design an attention sub-network based on the corresponding feature embedding $\mathbf{v}_i$ and $\mathbf{v}_j$. Specifically,  we first project $\mathbf{Z}=concat[\mathbf{v}_i,\mathbf{v}_j]\in \mathbb{R}^{1\times 2d}$ as input to obtain attention mask $\mathbf{I}$ as follows:

\begin{equation}
         \mathbf{I} = \phi_3\left(\phi_2\left(\mathbf{Z}\mathbf{W}_{2}\right )\mathbf{W}_{3}\right)\in \mathbb{R}^{1\times ml}
 \end{equation}
 where  $\mathbf{W}_2 \in \mathbb{R}^{ 2d\times s} $ and $\mathbf{W}_3 \in \mathbb{R}^{ s\times ml } $ are 
learning matrix of two MLP layers, $\phi_2\left(\cdot\right) $ is $ReLu \left(\cdot\right)$ and $\phi_3\left(\cdot\right) $ is identity function. Then, attention mask $\mathbf{I} $ is reshaped into $m $ groups:
  \begin{equation}
         \mathbf{I}^r = [\mathbf{I}_1,\mathbf{I}_2,......\mathbf{I}_m]\in \mathbb{R}^{m \times l}
 \end{equation}
where $\mathbf{I}_i \in \mathbb{R}^{1 \times l}$ is the corresponding attention map for $i $-th codeword embedding. We leverage attention matrix $ \mathbf{I}^r$ to recover the embedding by filtering out its information from each codeword, which mixes many cross feature's information. The procedure is computed as follows:
  \begin{equation}
  \begin{aligned} 
         \mathbf{C}_I^{(i,j)} & =\mathbf{I}^r \otimes \mathbf{C}^{(i,j)} \in \mathbb{R}^{m \times l}
         \\ &= [\mathbf{I}_1\otimes \mathbf{C}_1^{(i,j)},\mathbf{I}_2\otimes \mathbf{C}_2^{(i,j)},......,\mathbf{I}_m\otimes \mathbf{C}_m^{(i,j)}]
\end{aligned} 
 \end{equation}
 where  $\otimes$ is an element-wise multiplication between two vectors. The following procedure is similar to LMR and we flatten the attentive matrix $\mathbf{C}_I^{(i,j)}$ and map it into a new embedding space via MLP according to formula 11 to better restore cross feature's embedding $\mathbf{v}^{(i,j)}$. We call this method "Attentive Memory Restoring(AMR)" in this paper.

 We find AMR outperforms LMR when the codeword number is relatively small while LMR performs better if we continue to increase the codeword's number. This indicates 1-order feature embedding indeed helps recover cross feature's representation because a small codeword number means much more mixed chunk representations in one codeword. We will discuss this in Section 5.3.
 
\textbf{Feature Shrinking.}
In this section, we describe our proposed feature shrinking approaches. Cross features should be compressed in this stage due to the following two reasons: First, a large amount of cross feature's embedding brings noise because of its highly sparse property. Second, there are $f \times (f-1)$ new 2-order cross features in an input instance and the embedding layer will be too wide if we don't compress them, let alone 3-order or even higher order feature interactions.

In this stage, we aim to compress cross features according to the following rule: Given a feature set $\mathcal{E}_2^{i} = \left\{  \mathbf{x}^{(i,j)} \right\}_{1 \leqslant j \leqslant f}$ which contains all 2-order cross features derived by feature $ \mathbf{x}_i$, we leverage a weighted sum function $ \delta(\cdot)$ to project all corresponding embedding into $ \mathbf{v}_2^i \in \mathbb{R}^{1 \times d} $, where $ \mathbf{v}_2^i $ is a compressed feature embedding as the representation of all useful cross features derived by $ \mathbf{x}_i$. Formally, we have following function $ \delta(\cdot)$:
 \begin{equation}
         \mathbf{v}_2^i = \sum_{j=1}^{f}\mathbf{a}_j \mathbf{v}^{(i,j)}, j \neq i
 \end{equation}
Therefore, we need another attentive function $ \psi(\cdot)$ to evaluate the importance of each new feature derived by $ \mathbf{x}_i$. 

We leverage all the 1-order feature embedding to help select useful cross features. We think 1-order feature is not so sparse as high order features and contains more information, which helps reduce noisy cross feature's negative effect. Formally, we first concatenate embedding of all 1-order features as $\mathbf{V} = concat[\mathbf{v}_{1},\mathbf{v}_{2},...,\mathbf{v}_{f}]\in \mathbb{R}^{1\times fd}$. Then, $\mathbf{V} $ is input into an attention network to produce vector-wise attention score for each pairwise feature  in feature set $\mathcal{E}_2^{all} $ as follows:
\begin{equation}
         \mathbf{R} = \phi_5\left(\phi_4\left(\mathbf{V}\mathbf{W}_{4}\right )\mathbf{W}_{5}\right)\in \mathbb{R}^{1\times f(f-1)}
 \end{equation}
 where  $\mathbf{W}_4 \in \mathbb{R}^{ fd\times s} $ and $\mathbf{W}_5 \in \mathbb{R}^{ s\times f(f-1) } $ are 
parameter matrix of two MLP layers, $\phi_4\left(\cdot\right) $ is  $ReLu \left(\cdot\right)$,and $\phi_5\left(\cdot\right) $ is an identity function without non-linear transformation. Here  $\mathbf{R}\in \mathbb{R}^{1\times f(f-1)} $ is the attention map with size of $ f \times (f-1) $ , which contains weight for each feature in  set $\mathcal{E}_2^{all} $. Then, we can compress features according to formula 15 after obtaining $\mathbf{R}$. We call this approach "Global Attentive Shrinking(GAS)" in this paper.

Due to the feature shrinking, we can compress all 2-order features of an input instance from $ 1 \times f(f-1)d$ size layer into a thin embedding layer $ \mathbf{V_2}$ with size of $ 1\times fd$ :
\begin{equation}
        \mathbf{V_2} = concat[\mathbf{v}_{2}^1,\mathbf{v}_{2}^2,...,\mathbf{v}_{2}^f]\in \mathbb{R}^{1\times fd}
 \end{equation}

We can learn and memorize representation of high order features just as 2-order HCNet does, which also consists of the above-mentioned three phases. 
%

\subsection{MemoNet}
HCNet can be regarded as a general module to efficiently memorize pairwise or high order cross features. Therefore, we incorporate it into DNN model described in section 3 to boost its performance. We call the HCNet with DNN backbone "MemoNet" and the overall structure is shown in Figure 2.

As discussed in section 2.1, modern complex deep CTR models are inclined to design sub-network to capture cross features such as CIN in xDeepFM\cite{lian2018xdeepfm} and bi-linear interaction in FiBiNet\cite{HuangZZ19}. However, MemoNet use HCNet to memorize cross features with limited resources and the DNN part is only responsible for generalization of all features, including 1-order, 2-order and high order features. This kind of clear division of labour for each component in MemoNet greatly reduces the learning difficulty and boosts model performance. Besides, as the main memory in HCNet, size of multi-hash codebook can be gradually enlarged to store more useful knowledge in cross features and further boost the model's performance just as big models in NLP show. We will verify these through our experiments.

Though MemoNet leverages DNN as backbone of the model, HCNet can be used as a general plug-in module in other CTR models. We explore the compatibility of our proposed HCNet and integrate xDeepFM, DCN v2 and MaskNet with HCNet in Section 5.5.  

\subsection{Key Interaction Field}
In this section, we concisely discuss the time complexity of HCNet and propose an approach to greatly increase HCNet's efficiency.

Suppose we have $\mathbf{f}$ feature fields and 2-order HCNet needs to compute $\mathbf{f \times (f-1)/2}$ cross features through the proposed method, which means high time complexity if $\mathbf{f}$ is a large number. To tackle this problem and increase the model's efficiency, we propose the following method to greatly reduce the time complexity of HCNet from $\mathbf{f \times (f-1)/2}$ to $\mathbf{m \times (f-1)/2}$. Here  $\mathbf{m}$ is a much smaller number than feature fields number $\mathbf{f}$.  

Given dataset, we find only a small fraction of fields are important for feature interaction, which contributes majority of the performance gain in HCNet. We name this kind of field "key interaction field(KIF)" and the field number is $\mathbf{m}$. Therefore, we can only use features in these KIFs to interact with any other feature in input instance, which will greatly reduce the combination number of cross features in HCNet from  $\mathbf{f \times(f-1)/2}$ to $\mathbf{m\times(f-1)/2}$ and increase HCNet's efficiency. In order to identify these KIFs, we propose to use score function  $ \mathbf{s}_{F_k}(\cdot)$  to rank field importance of feature interaction, and the top scored fields can be regarded as KIFs.

In this paper, we propose two approaches to compute the field's importance. A straightforward method is to rank field according to the feature number it contains. More features it contains, the more important it is. Formally, we have the following field importance score for th-$\mathbf{k}$ field:
 \begin{equation}
         \mathbf{s}_{F_k} = \Vert F_k \Vert
 \end{equation}
where $\Vert F_k \Vert$ means the number of features which belong to field $F_k$. We call this method 'feature number ranking'(FNR).

The second method is named "field attention ranking(FAR)" because we can compute attention score for one field by accumulating the GAS attention score of features belonging to this field. The higher the attention score one field has, the more important it is. Specifically, after training HCNet, we can leverage a small validation dataset to find out which field is vital by accumulating the attention score of the feature in one field. We can compute the field importance score $\mathbf{s}_{F_k}$ for the-$\mathbf{k}$ field as follows:
 \begin{equation}
         \mathbf{s}_{F_k} = \sum_{i=1}^{n}\sum_{j=1}^{f}\mathbf{a}_{(k,j)}^i , j \neq k
 \end{equation}
where $n$ denotes the instance number of validation set, $f$ is field number and $\mathbf{a}_{(k,j)}^i$ is the attention score of th-$(k,j)$ cross feature in th-$i$ instance computed by GAS.

\section{Experimental Results}

\subsection{Experiment Setup}

\textbf{Datasets.}
We evaluate our model on three real-world datasets while AUC (Area Under ROC) and Logloss (binary cross-entropy loss) are used as the evaluation metrics in our experiments.

\begin{table}[h]
\centering
\caption{Statistics of the Evaluation Datasets}
\begin{tabular}{lccc}
\toprule
Datasets  & \#Instances & \#fields & \#features \\
\midrule
Avazu       & 40.4M  & 23 & 9.4M     \\
KDD12     & 194.6M  & 13 & 54.6M \\
Criteo       & 45M  & 39 & 33.7M     \\
\bottomrule
\end{tabular}
\label{tab:datasets}
\end{table}

\textbf{(1) Avazu\footnote{Avazu \url{http://www.kaggle.com/c/avazu-ctr-prediction}} Dataset:} The Avazu dataset consists of several days of ad click- through data which is ordered chronologically. For each click data, there are 23 fields which indicate elements of a single ad impression.
\textbf{(2) KDD12\footnote{KDD12 \url{https://www.kaggle.com/c/kddcup2012-track2}} Dataset:} KDD12 dataset aims to predict the click-through rate of ads. There are 13 fields spanning from user id to ad position for a clicked data.
\textbf{(3) Criteo\footnote{Criteo \url{http://labs.criteo.com/downloads/download-terabyte-click-logs/}} Dataset:} Criteo dataset is widely used in many CTR model evaluation. There are 26 anonymous categorical fields and 13 continuous feature fields in Criteo dataset.

We randomly split instances by 8:1:1 for training, validation, and testing while Table \ref{tab:datasets} lists the statistics of the evaluation datasets. 

\textbf{Models for Comparison.}
We compare performances of FM\cite{rendle2010factorization}, Wide\&Deep\cite{cheng2016wide}, DNN\cite{zhang2016deep}, DeepFM\cite{guo2017deepfm}, DCN\cite{wang2017deep}, AutoInt\cite{song2019autoint}, DCN V2\cite{WangSCJLHC21}, xDeepFM\cite{lian2018xdeepfm}, FiBiNet\cite{HuangZZ19}, CAN\cite{zhou2020can}
and MaskNet\cite{masknet} models as baselines and all of which are discussed in Section 2. Results of some models such as LR\cite{he2014practical}, FwFM\cite{pan2018field}, NFM\cite{nfm}, AFM\cite{xiao2017attentional} and CCPM\cite{ccpm} are not presented in this paper, because more recent models like FiBiNET\cite{HuangZZ19} and DCN-V2\cite{WangSCJLHC21} have  outperformed these methods significantly as experiments in FuxiCTR\cite{jieming-2009-05794} shows.

\textbf{Implementation Details.}
For the optimization method, we use the Adam\cite{adam} with a mini-batch size of $1024$.  We make the dimension of 1-order feature embedding for all models to be a fixed value of $10$ for Criteo,  $50$ for Avazu, and $10$ for KDD12 dataset.  For models with DNN part, the depth of hidden layers is set to $3$, the number of neurons per layer is $400$, and all activation functions are ReLU. In HCNet, unless otherwise specified, we use 2 hash functions and 1M codewords(0.5M codewords on KDD12 dataset) as our default setting and keep the size of codeword embedding the same with 1-order feature embedding. For other models, we take the optimal settings from the original papers.

\subsection{Performance Comparison}

\begin{table*}
\centering
\caption{Overall performance of different models(MC means million codewords and  $x$H denotes hash function number$=x$)}
\begin{tabular}{c|l|lllllllll}
\toprule
 \multicolumn{2}{c}{} & \multicolumn{3}{c}{\textbf{Avazu}}  & \multicolumn{3}{c}{\textbf{KDD12} } & \multicolumn{3}{c}{\textbf{Criteo}} \\
\midrule
\multicolumn{2}{c}{Model} & AUC(\%) $\uparrow$ &  Logloss $\downarrow$ &Paras. & AUC(\%) $\uparrow$ & Logloss $\downarrow$ &Paras. & AUC(\%) $\uparrow$ & Logloss $\downarrow$ &Paras.   \\
\midrule
Shallow &FM       & 78.17 &0.3809 &79M  & 77.65 &0.1583 &60M & 78.97  & 0.4607 &11M \\
\midrule
Mining &Wide\&Deep      & 78.49 &0.3788 &79M  & 79.33 &0.1542 &60M & 80.53  & 0.4462 &11M   \\
   &AutoFIS       & 78.79 &0.3745 &79M  & 79.54 &0.1532 &60M & 80.77  & 0.4432 &11M   \\
\midrule
   &DeepFM       & 78.64 &0.3769 &79M  & 79.40 &0.1538 &60M & 80.58  & 0.4457 &11M   \\
 &xDeepFM       & 78.88 &0.3747 &81M  & 79.51 &0.1534 &62M & 80.64  & 0.4450 &15M   \\
 Modeling &DCN       & 78.68 &0.3767 &78M  & 79.58 &0.1531 &55M & 80.73  & 0.4441 &10M   \\
 &FiBiNet       & \underline{79.12} &\underline{0.3730} &87M  & 79.52 &0.1533 &61M & 80.73  & 0.4441 &17M   \\
  &AutoInt+       & 78.62 &0.3758 &78M  & 79.69 &0.1529 &55M & 80.78  & 0.4438 &10M   \\
  &MaskNet       & 78.94 &0.3740 &85M  & \underline{79.89} &\underline{0.1521} &56M & \underline{81.07}  & \underline{0.4410} &13M   \\
   &DCN v2       & 78.98 &0.3738 &81M  & 79.66 &0.1530 &55M & 80.88  & 0.4430 &11M   \\
    &CAN       & 79.02 &0.3737 &223M  & 79.71 &0.1530 &569M & 80.81  & 0.4440 &105M   \\
\midrule
\midrule
Backbone&DNN       & 78.67 &0.3756 &78M  & 79.54 &0.1533 &55M & 80.73  & 0.4440 &10M   \\
\midrule
&  MemoNet-L & \textbf{79.58/9H} & \textbf{0.3704} &128M  & \textbf{80.60/7H} &\textbf{0.1508} &60M/0.5MC & 81.28/3H  & 0.4390 &20M    \\
\textbf{Ours}&  \quad \emph{Improve.} & +0.91 & -0.0052 &+50M  & +1.06 &-0.0025 &+5M & +0.55  & -0.0050 &+10M    \\
(Small/1MC)&  MemoNet-A & 79.52/8H & 0.3707 &128M  & 80.52/8H &0.1516 &60M/0.5MC & \textbf{81.30/3H}  & \textbf{0.4387} &20M    \\
& \quad \emph{ Improve.} & +0.85 & -0.0049 &+50M  & +0.98 &-0.0017 &+5M & +0.57  & -0.0053 &+10M    \\
 \midrule 
 &   MemoNet-L & \textbf{79.87/9H} & \textbf{0.3684} &578M  & - &- &- & \textbf{81.39/4H} & \textbf{0.4380} &  110M  \\
\textbf{Ours}& \quad \emph{ Improve.} & +1.20 & -0.0072 &+500M  & - &- &- & +0.66  & -0.0060 &+100M    \\
(Big/10MC)&  MemoNet-A & 79.73/9H &0.3693 &578M  & - &- &- & 81.38/3H  & 0.4381 &110M    \\
&  \quad \emph {Improve.} & +1.06 & -0.0063 &+500M  & -&- &- & +0.65  & -0.0059 &+100M    \\
   
\bottomrule
\end{tabular}
\label{tab:overalperformance}
\end{table*}

Table \ref{tab:overalperformance} shows the performance of different baselines and MemoNet. For MemoNet, two groups of experiments are designed: the first group is small MemoNet which has a codebook with 1 million codewords. The second group is big MemoNet containing a codebook with 10 million codewords. We fix other settings and tune the number of hash functions for the best performance model.

The experiments for MemoNet and the baseline models are repeated 5 times by changing the random seeds and the averaged results are reported for a fair comparison. The best results are in bold, and the best baseline results are underlined.
\begin{figure*}[h]
  \centering
  \includegraphics[width=0.8\linewidth]{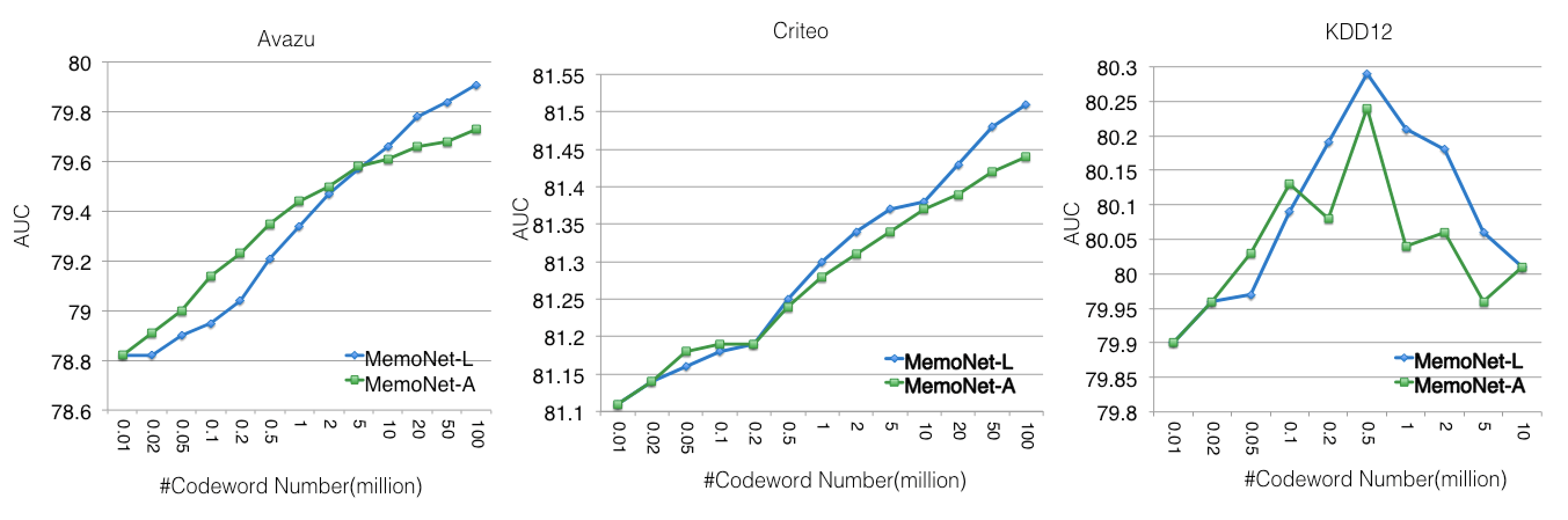}
  \caption{Effect of different codeword number in HCNet}
  \Description{}
\end{figure*}
From the experimental results, we mainly have the following observations:

(1) Compared with other baselines, two small MemoNets(MemoNet-L with linear memory restoring and MemoNet-A with attentive memory restoring approach) both achieve the best results on all benchmarks and outperform other SOTA models with a large margin. The best performance small MemoNet models show a huge superiority in performance and outperform DNN backbone by +0.91,  +1.06, and +0.57 absolute value in terms of AUC on three datasets, respectively. This indicates that a specifically designed  memorizing mechanism for cross features indeed makes DNN part focus on feature generalization and reduces the learning difficulty. On the other hand, HCNet respectively contains 50M, 5M and 10M parameters to memorize cross features on three datasets, which is applicable in most real life CTR scenes. The experimental results demonstrate the feasibility of memorizing all cross features with limited resources to greatly boost model performance.

(2) As for the big MemoNet, which contains 500M and 100M parameters in HCNet on Avazu and Criteo datasets respectively, we find the model performance continues to increase compared with small MemoNet. Besides, best performance MemoNets outperform best baselines by +0.75, +0.71, and +0.32 absolute value in terms of AUC on three datasets, respectively. It is worth noting that, when comparing with the best baseline, an improvement of AUC at 0.1 level is rather hard in real-life CTR tasks. 

Compared with the backbone DNN model, small MemoNet's model size increases 65\%, 9\% and 100\% on three datasets respectively. \textbf{But larger model size is not the reason why MemoNet outperforms baselines.} Increasing the feature embedding size or DNN layer indeed have an influence on model performance. However, we have set the optimal feature embedding size for most baselines in our experiments and performances will drop if we continue to increase model size because of over-fitting.  We also conduct experiments on DNN backbone model by increasing MLP layer number to make the model bigger. However, the gain is rather small and AUC performance improvement is between 0.1 and 0.2 compared with 3 layer MLP model, which is consistent with the hyper-parameter experiment results shown by many published papers\cite{guo2017deepfm}\cite{HuangZZ19}\cite{lian2018xdeepfm}. This indicates that larger SOTA baselines don't lead to much better model performance because of the parameter over-fitting.

We think the reason why MemoNet outperforms modeling based CTR models such as CAN and DCN v2 is because memorizing cross features by a specifically designed  memorizing mechanism keeps much more useful information than modeling cross features through sub-network. For example, CAN tries to learn a micro-MLP model for each feature to capture cross features, which leads to much more model parameter than MemoNet does, the performance of CAN still obviously underperforms MemoNet. On the other hand, MemoNet tries to learn and memorize all cross features' representations, including a large amount of long-tail useful cross features, which can explain why MomoNet outperforms mining based CTR models such as Wide \& Deep and AutoFIS because these methods can only select few most useful cross features, manually or automatically.
\subsection{Effect of Codeword Number}
We conduct extensive experiments on three datasets to explore the effects of various numbers of codewords in the codebook of HCNet on performance and show the different characteristics of MemoNet-L and MemoNet-A models.

We gradually increase codeword number from 0.01M to 100M on Avazu and Criteo(the max codeword number is set to 10M on KDD12 because 0.5M is the optimal number) to watch the performance change. Figure 3 reports the experimental results in MemoNet-L  and MemoNet-A and we have the following findings:

Finding 1:\textbf{ MemoNet achieves large language model's scaling law by absorbing more knowledge from data when we increase model parameters through memorizing.} We can see from Figure 3 that both models'(MemoNet-L and MemoNet-A) performances continue to increase when we enlarge the codeword number on Avazu and Criteo datasets. The procedure is stopped when the max codeword number reaches 100M (Avazu with AUC of 79.91\% and Criteo with AUC of 81.52\%) because of our limited hardware resources, and it seems the performance will further increase if we continue to enlarge codeword number. It demonstrates that MemoNet indeed obtains the big model's ability of absorbing more knowledge from data when we increase model parameters through memorizing. We argue that MemoNet is the first CTR model to have this kind of ability as big models in NLP show.  However, the optimal codeword number is 0.5M on KDD12 dataset. This indicates KDD12 dataset has much less useful cross features than Avazu and Criteo, and small HCNet can effectively memorize all effective cross features with rather limited resources. If the dataset contains a large amount of long-tail useful cross features such as Avazu, we can just enlarge the codeword number of codebook in HCNet to sustainably obtain performance gains. 
 
Finding 2:\textbf{ MemoNet-A outperforms MemoNet-L when the codebook is small.} It can be seen from Figure 3 that MemoNet-A performs better than MemoNet-L model when the codeword number is relatively small and it will go into reverse once the codeword number reaches a specific number. On KDD12 and Criteo, the turning point number is 0.2M and it's 5M on Avazu dataset. This denotes the attentive memory restoring helps recover cross feature's representation because there are more cross features sharing the same codeword embedding when the codeword number is small.
\begin{figure}[h]
  \centering
  \includegraphics[width=0.7\linewidth]{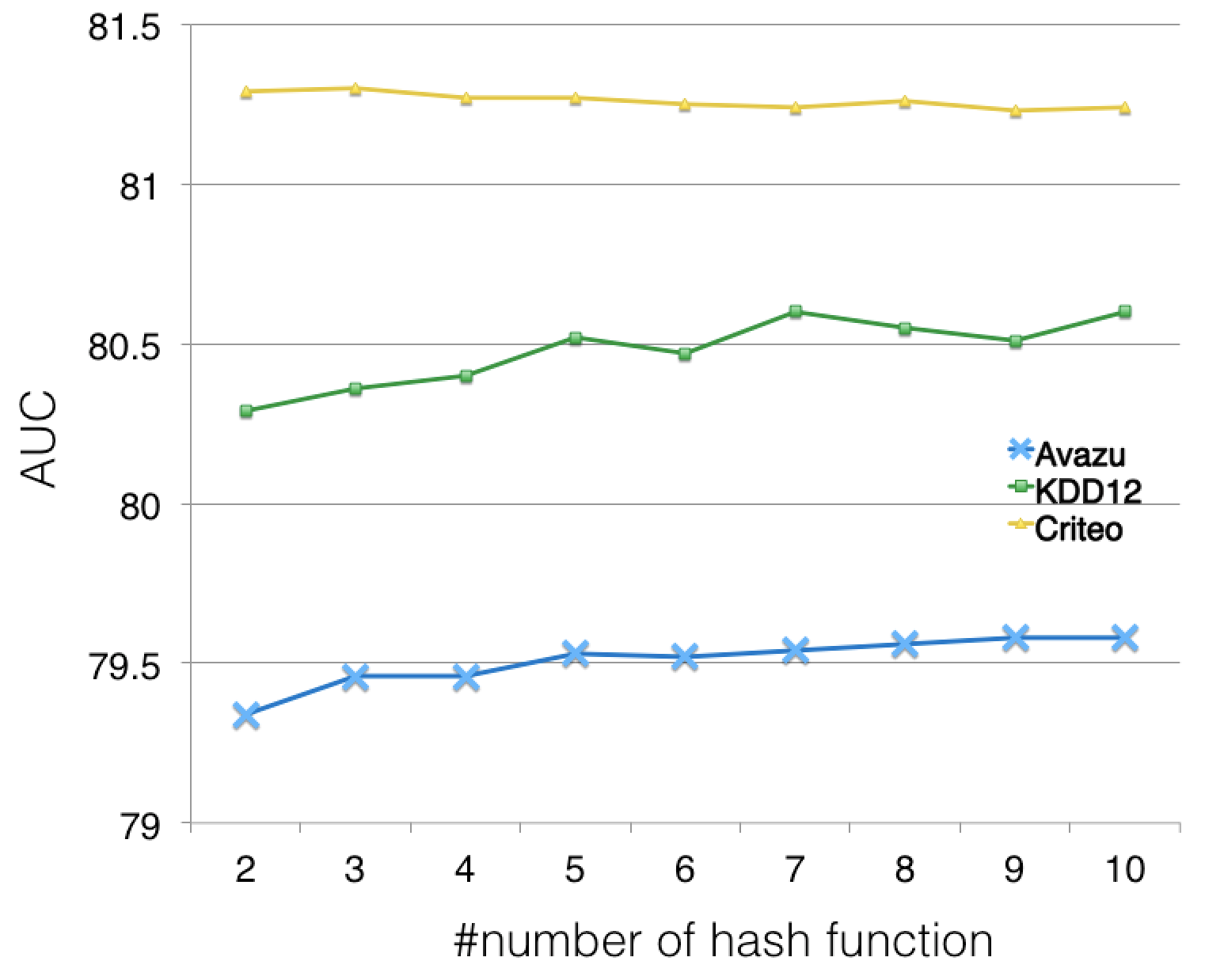}
  \caption{Effect of different hash numbers in HCNet}
  \Description{}
\end{figure}
\subsection{Effect of Hash Function Number}
This section studies performance sensitivity when setting different numbers of hash functions in HCNet, as shown in Figure 4. We find the performance sensitivity of hash function number depends on datasets. Firstly, we observe that the performance of MemoNet-L on Criteo fluctuates within a limited range when changing hash function numbers. However, model performance on Avazu will increase continuously with the increase of hash function number. As for the model performance on KDD12, it increases continuously when the hash function number is less than 5 and begins to fluctuate after that. The possible reason is that Avazu dataset contains much more useful long-tail cross features compared with the other two datasets.

\begin{table}
\centering
\caption{Performance Gains (AUC) of SOTA models after incorporating 2-order HCNet}
\begin{tabular}{l|cccccc}
\toprule
  & \multicolumn{1}{c}{\textbf{Avazu}}  & \multicolumn{1}{c}{\textbf{KDD12} } & \multicolumn{1}{c}{\textbf{Criteo}} \\
\midrule
Model & AUC(\%)   & AUC(\%)   & AUC(\%)    \\
\midrule
xDeepFM       & 78.88    &  79.51  & 80.64    \\
 xDeepFM${_{HCNet}} $       & 79.49    & 80.34  & 81.22    \\
\quad \quad$\Delta$           & \textbf{+0.61}  &\textbf{+0.83} &\textbf{+0.58}  \\
 \midrule 
 DCN v2       & 78.98    &  79.66  & 80.88    \\
 DCN v2${_{HCNet}} $       & 79.42    & 80.37  & 81.36    \\
\quad \quad$\Delta$           & \textbf{+0.44}  &\textbf{+0.71} &\textbf{+0.48}  \\
 \midrule 
MaskNet       & 78.94    &  79.89  & 81.07    \\
\textbf{MaskNet${_{HCNet}} $  }     & \underline{79.57}   & \underline{80.50}  & \underline{81.40}    \\
\quad\quad $\Delta$           & \textbf{+0.63}  &\textbf{+0.61} &\textbf{+0.33}  \\
 \midrule 
 \midrule 
DNN       & 78.67    &  79.54  & 80.73    \\
MemoNet       & 79.44    & 80.29  & 81.30    \\
\quad\quad $\Delta$           & \textbf{+0.77}  &\textbf{+0.75} &\textbf{+0.57}  \\

\bottomrule
\end{tabular}
\label{tab:performancegains}
\end{table}

\subsection{Plugging HCNet into SOTA Models}
As a concise and effective component, we argue that HCNet can be easily incorporated into any deep CTR model to boost performance. To verify this, extensive experiments are conducted on three strong baselines: xDeepFM, DCN v2 and MaskNet. Table 3 shows the results which demonstrate that the enhanced models significantly outperform the original methods with a large margin. It demonstrates that the proposed solution can effectively make the deep models learn cross features easily through memorizing.

\subsection{Effect of Methods to Identify KIFs}
\begin{figure}[h]
  \centering
  \includegraphics[width=\linewidth]{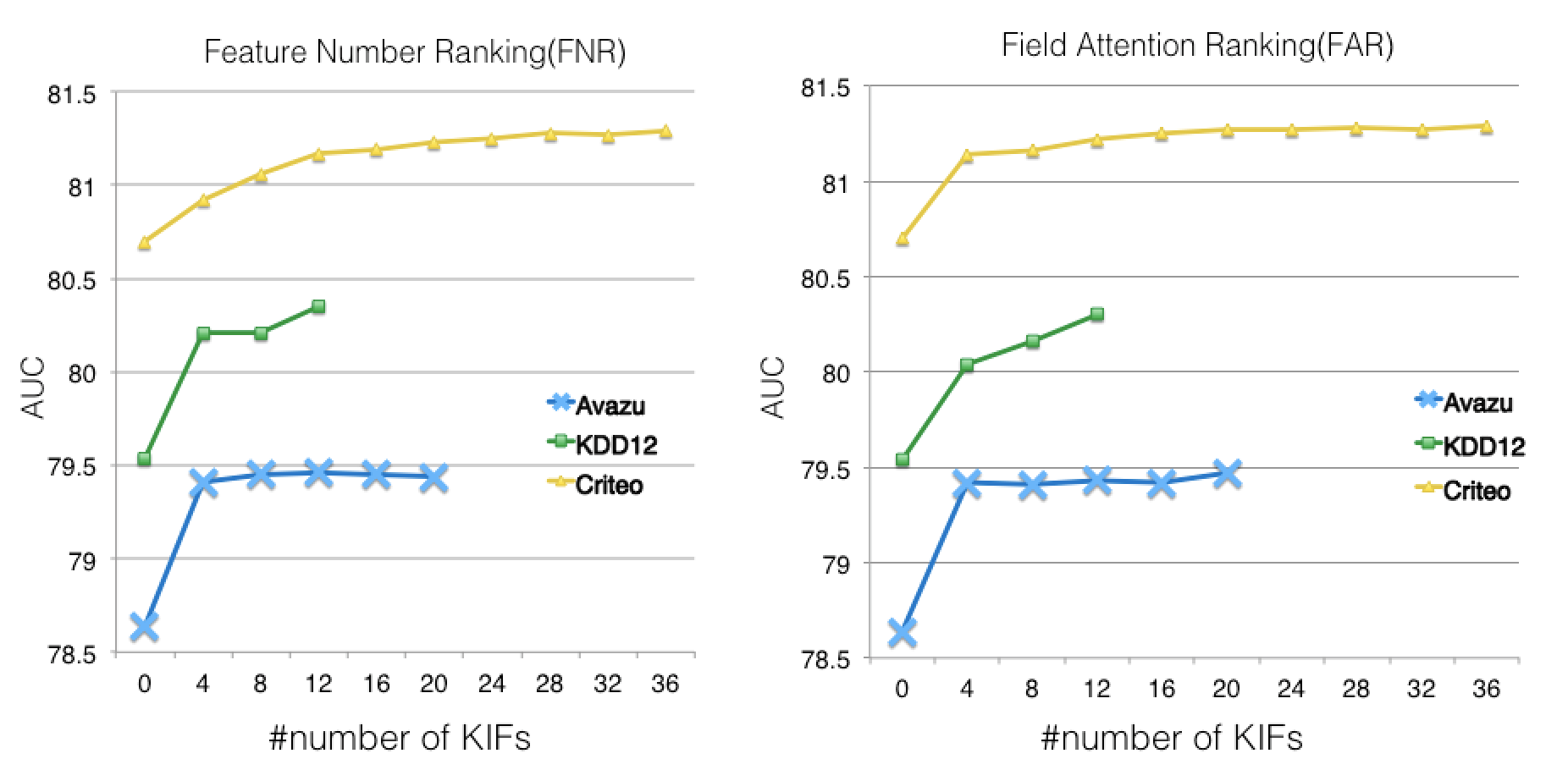}
  \caption{Effect of FNR and FAR}
  \Description{}
\end{figure}
In this section, two groups of experiments are conducted to show the effectiveness of our proposed "feature number ranking(FNR)" and "field attention ranking(FAR)" methods to identify the "key interaction field" in order to greatly reduce the feature interaction number in HCNet. First, we rank fields according to FNR and FAR scores respectively. Then, top scored fields are gradually added into HCNet to watch the change of model performance. For all three datasets, we select four fields to be added each time and Figure 5 shows the results. We can see from the results that: Both the FNR and FAR methods can successfully identify the most beneficial fields. Take FAR as an example, if we select top 4 ranked fields as KIFs, the performance gain ratio of HCNet will be 93.90\%, 74.58\%, and 65.79\% on Avazu, Criteo and KDD12 datasets, respectively. Here performance gain ratio denotes the ratio between the performance gain of HCNet with the top 4 KIFs compared with DNN and that gain of HCNet using all fields as KIFs. This shows we can leverage a little fraction of fields to obtain the majority of performance gain, which will greatly reduce time complexity of HCNet.

\subsection{Online Deployment and A/B Test}
We have deployed the proposed 2-order HCNet module as a plug-in component on a large-scale recommendation system in one of the largest social-media platforms for feed recommendation. The strict online A/B test experiments were conducted for a duration of 14 days, from Dec 10, 2022, to Dec 23, 2022. The baseline is the last deployed production model, a DNN-based deep model. We observed consistently significant online A/B testing performance gains with 9.73\% improvement on CTR and 6.86\% improvement on Watch Time. We adopt FAR method to use only 4 KIFs among 96 feature fields and the online latency increased 10.33\%. These online experiment results demonstrate the efficiency and effectiveness of our proposed HCNet model.
\subsection{Effect of Higher Order HCNet}
\begin{figure}[h]
  \centering
   \includegraphics[width=0.7\linewidth]{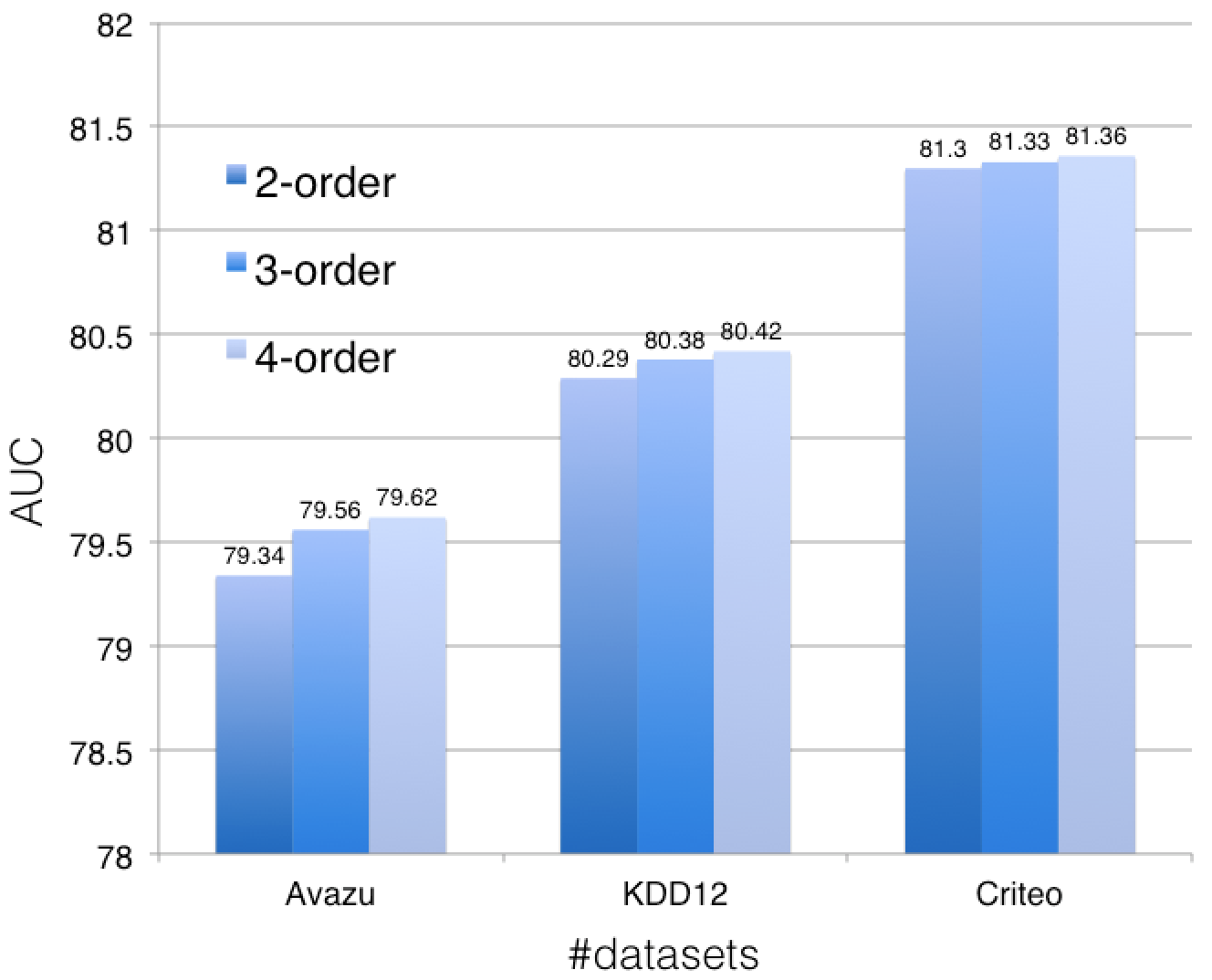}
  \caption{Model performance of high order HCNet}
  \Description{}
\end{figure}
In this section, we evaluate how the performance changes when we plug high order HCNets into the MemoNet-L model. Note we allocate new storage resources with size of 1M codewords in codebook of high order HCNet. Their performances on three benchmarks are presented in Figure 6, from which we can see that: model performance consistently increases when we add higher order HCNet into MemoNet-L on all three datasets. This indicates high order features are beneficial to boost model performance. 

\section{Conclusion and Future Work}
We propose HCNet as a memory mechanism for efficiently learning and memorizing representations of cross features in CTR tasks. We also propose  MemoNet model by combining HCNet with a DNN backbone. Extensive experimental results on three public datasets show that MemoNet outperforms SOTA models. In future work, we will explore more complex codebook structures.

\bibliographystyle{ACM-Reference-Format}
\balance
\bibliography{sample-base}



\end{document}